\documentclass[namedreferences]{solarphysics}

\usepackage[hyperref,optionalrh,solaromanenum]{spr-sola-addons} % For Solar Physics 
\usepackage{graphicx}        % For eps figures, newer & more powerfull
\usepackage{color}           % For color text: \color command
\usepackage{breakurl}        % For breaking URLs easily trough lines
\newcommand\arcsec{\mbox{$^{\prime\prime}$}}%
\newcommand\ion[2]{#1$\;${\small {#2}}\relax}% 

\chardef\us=`\_

\begin{document}

\begin{article}
\begin{opening}

\title{On the Evolution of a Sub-C Class Flare: A Showcase for the Capabilities of the Revamped Catania Solar Telescope}

\author[addressref=aff1,corref,email={paolo.romano@inaf.it}]{\inits{P.}\fnm{Paolo}~\lnm{Romano}\orcid{0000-0001-7066-6674}}
\author[addressref=aff1,email={salvatore.guglielmino@inaf.it}]{\inits{S.L.}\fnm{Salvo~L.}~\lnm{Guglielmino}\orcid{0000-0002-1837-2262}}
\author[addressref=aff1]{\inits{P.}\fnm{Pierfrancesco}~\lnm{Costa}}
\author[addressref=aff1]{\inits{M.}\fnm{Mariachiara}~\lnm{Falco}}
\author[addressref=aff1]{\inits{S.T.}\fnm{Salvatore}~\lnm{Buttaccio}}
\author[addressref=aff1]{\inits{A.}\fnm{Alessandro}~\lnm{Costa}}
\author[addressref=aff1]{\inits{E.}\fnm{Eugenio}~\lnm{Martinetti}}
\author[addressref=aff1]{\inits{G.}\fnm{Giovanni}~\lnm{Occhipinti}}
\author[addressref=aff1]{\inits{D.}\fnm{Daniele}~\lnm{Spadaro}}
\author[addressref=aff1]{\inits{R.}\fnm{Rita}~\lnm{Ventura}}
\author[addressref=aff2]{\inits{G.E.}\fnm{Giuseppe~E.}~\lnm{Capuano}}
\author[addressref=aff2]{\inits{F.}\fnm{Francesca}~\lnm{Zuccarello}}

%\author{\inits{}\fnm{}~\lnm{}\orcid{}}
%   NOTE:  Just one corresponding author [corref]
%   \institute{$^{1}$ First affiliation
%                     email: \url{e.mail-a} email: \url{e.mail-b}\\ 
%              $^{2}$ Second affiliation
%                     email: \url{e.mail-c} \\
%             \textit{}
\address[id=aff1]{INAF - Osservatorio Astrofisico di Catania, Via S.~Sofia 78, I-95123 Catania, Italy}
\address[id=aff2]{Dipartimento di Fisica e Astronomia ``Ettore Majorana'' -- Sezione Astrofisica, Universit\`a degli Studi di Catania, Via S.~Sofia 78, I-95123 Catania, Italy}

\runningauthor{Romano et al.}
\runningtitle{Capabilities of the Revamped Catania Solar Telescope}

\begin{abstract}
Solar flares are occasionally responsible for severe space weather events, which can affect space-borne and ground-based infrastructures, endangering anthropic technological activities and even human health and safety. Thus, an essential activity in the framework of space weather monitoring is devoted to the observation of the activity level on the Sun. 

In this context, the acquisition system of the Catania Solar Telescope has been recently upgraded in order to improve its contribution to the European Space Agency (ESA) -- Space Weather Service Network through the ESA Portal, which represents the main asset for space weather in Europe. 

Here, we describe the hardware and software upgrades of the Catania Solar Telescope and the main data products provided by this facility, which include full-disk images of the photosphere and chromosphere, together with a detailed characterization of sunspot groups. As a showcase of the observational capabilities of the revamped Catania Solar Telescope, we report the analysis of a B5.4 class flare that occurred on 7 December 2020, simultaneously observed by the Interface Region Imaging Spectrograph and the Solar Dynamics Observatory satellites.

%\textcolor{red}{Lorem ipsum dolor sit amet, consectetur adipiscing elit, sed do eiusmod tempor incididunt ut labore et dolore magna aliqua. Ut enim ad minim veniam, quis nostrud exercitation ullamco laboris nisi ut aliquip ex ea commodo consequat. Duis aute irure dolor in reprehenderit in voluptate velit esse cillum dolore eu fugiat nulla pariatur. Excepteur sint occaecat cupidatat non proident, sunt in culpa qui officia deserunt mollit anim id est laborum.}

\end{abstract}
\end{opening}
%-------------------------------------------------

\section{Introduction}
	\label{introduction} 

Solar eruptions are the most prominent manifestations of the magnetic activity of the Sun, involving the entire heliosphere with a potential impact on the Earth \citep[e.g.][]{Patsourakos16,Piersanti17}. Their effects on the near-Earth environment and anthropic activities are increasingly involving socio-economical interests, taking into account that solar eruptions may provoke severe damages to technological systems. Therefore, many recent efforts of the scientific community have been addressed to forecast eruptive events with enough advance to prevent or mitigate their impact on human activities \citep{Schwenn06}.

Eruptive phenomena are usually associated with flares, which are sudden energy release events that emit radiation across the entire electromagnetic spectrum, as well as accelerated particles \citep[see][for a review]{Benz17}. Like eruptions, flares occur because of the release of free magnetic energy, which is converted into heat and kinetic energy through magnetic reconnection \citep{KP76,Moore01}. This leads to the observed brightness enhancements that can be easily detected also at visible wavelengths, such as the H$\alpha$ line, centered at $\lambda=656.28$~nm, which is one of the most popular lines for studying the solar chromosphere \citep[see][and references therein]{Leenarts12}. The flaring chromosphere is characterized by strong H$\alpha$ emission, related to the impact of energetic particles accelerated at the reconnection site colliding  in the chromospheric plasma. H$\alpha$ emission often appears as two bright ribbons, separating from each other \citep[see, e.g.,][]{Fletcher11}. However, considering the three-dimensional nature of magnetic reconnection \citep{Aulanier12,Aulanier13}, the presence of a more complex topology in the reconnection site may result in flares with extra ribbons or even circular ribbons \citep[e.g.][]{Masson09,Guglielmino16,Romano17}. 

Although the trigger of flares is thought to be located in the upper layers of the solar atmosphere, with the coronal magnetic field playing a leading role, most of the flare forecasting methods are based on photospheric observations of the active regions (ARs) where flares occur. Indeed, the magnetic field configuration suitable for the occurrence of eruptive events is mainly driven by the photospheric evolution of the ARs, determined by the emergence of new magnetic flux from the convection zone into the solar atmosphere and by the rearrangement of the coronal field due to the horizontal photospheric displacements of the field line footpoints \citep[e.g.][]{Romano07,Romano15,Romano18}, leading to magnetic helicity changes \citep[e.g.][]{Zuccarello21}. Thus, many approaches to flare forecasting produced in the last decades rely on the determination of certain photospheric parameters, such as the total unsigned magnetic flux together with the length of the magnetic polarity inversion line (PIL) characterized by strong line-of-sight field gradient and the total magnetic energy dissipation \citep{Yuan10}, or with the free magnetic energy obtained from the line-of-sight field gradient of the PILs \citep{Falconer08}. More recently, \citet{Korsos15} proposed the introduction of the weighted horizontal magnetic gradient ($WG_{M}$) to forecast flares based on the magnetic gradient among all spots within an appropriately defined region close to the PIL \citep[see also][]{Korsos16,Korsos19}.

Other forecasting methods are performed on a statistical base. \citet{Bloomfield12} proposed a method that uses the McIntosh group classification of sunspot groups (SGs) observed in the photosphere and assumed that flares are Poisson-distributed processes; they showed that Poisson probabilities perform comparably to other more complex prediction methods. The extensive comparison among several flare forecasting methods performed by \citet{Barnes16} suggested that it may be possible to obtain the best prediction by combining a method, which characterizes an AR by one or more parameters, and uses a statistical technique. In this perspective, \citet{Falco19} developed a method which is based mainly on the Zurich classification of the SGs observed at photospheric level and assumes the Poisson statistics for the flare occurrence. This method is able to provide an estimation of the capability to host flares of a specified energy range for an AR characterized by a particular configuration, size and fragmentation. 

In this article, we present the capabilities of the revamped Catania Solar Telescope. This facility, built in the 1960s, is now able to provide almost simultaneous full-disk observations of the chromosphere and photosphere in the H$\alpha$ line and in the nearby continuum with high cadence, down to 1~s, with an angular resolution of 2{\mbox{$^{\prime\prime}$}}. Such observations are an effective tool for a synoptic monitoring of the flaring activity of the chromosphere, as well as for characterizing several photospheric parameters of the ARs where flares occur.

The article is organized as follows. In Section~2 we describe the upgrade of the Catania Solar Telescope. Section~3 reports on the observation of a B5-class flare, simultaneously observed by the Catania Solar Telescope and by the Solar Dynamics Observatory (SDO) and the Interface Region Imaging Spectrograph (IRIS) satellites, as a showcase for illustrating the capabilities of the revamped facility. In Section~4 we draw our conclusions.

\section{The Catania Solar Telescope}
\label{Catania} 
The observations of the solar photosphere and chromosphere are carried out at the Istituto Nazionale di Astrofisica (National Institute for Astrophysics) -- Catania Astrophysical Observatory (INAF-OACT), in collaboration with the University of Catania, by means of a telescope equipped with two refractors with a diameter of 150 mm and a focal length of 2230 mm and 2300 mm, respectively. The first one is used to make drawings of sunspot groups and pores from visual observations, the second one feeds a Zeiss Lyot filter (bandwidth of 0.025 or 0.050 nm, tunable filter range $\pm$0.1 nm), which is used to take digital full-disk images of the photosphere in the continuum of the H$\alpha$ line at 656.78 nm and of the chromosphere in the center of the H$\alpha$ line at 656.28 nm

On a daily basis, when the weather conditions allow, a drawing of the projected Sun is performed in order to determine some properties of the sunspot groups visible in the photosphere for each of them: heliographic latitude and longitude of the baricenter, number of sunspots and pores, projected area in tens of millionths of the solar hemisphere, type of penumbra of the main sunspot of the group, relative importance between the leading spot and density of the sunspot population \citep[see][]{Ternullo06}, and group type according to the Zurich classification \citep{Falco19}. Given the characteristics of the instrument and average seeing conditions, the number of pores that can be detected by visual inspection of the projected photospheric image results to be greater than that retrieved by the digitized images, thanks to the better resolution of the former option. The photospheric data acquired at the OACT are distributed to international Solar Data Centers, like the Solar Influences Data Center (SIDC) in Brussels and the World Data Center for the Sunspot Index (NOAA, Boulder). 

When the weather conditions permit, digital images are also acquired in the center of the H$\alpha$ line and in the nearby continuum by the Lyot filter. The OACT contributes with its H$\alpha$ images to the Global High Resolution H$\alpha$ Network. The digital images of the Sun are also provided, on request, to observers carrying out observational campaigns using high-resolution solar telescopes.

The sunspot group characterization, the photospheric, and the chromospheric data are published in near real time on the portal of the ESA Space Situational Awareness Programme (https://swe.ssa.esa.int/solar-weather).

\begin{figure}   
	\centerline{\includegraphics[width=\textwidth,trim= 0 0 0 0,clip]{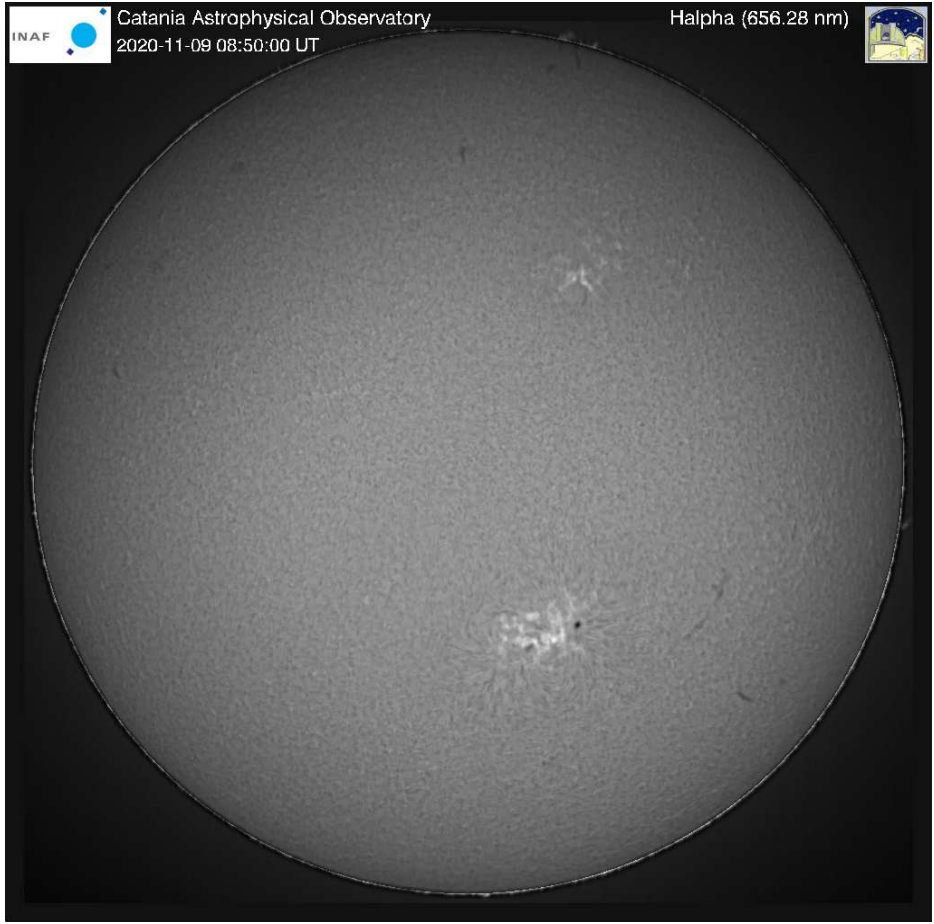}
	}
	\caption{Example of full-disk H$\alpha$ image taken by the Catania Solar Telescope during good seeing conditions and corrected by the flat field. Here and in the following figures, solar north is at the top, and west is to the right.
	}
	\label{fig:example}
\end{figure}

\subsection{Hardware Upgrade}
\label{HWup} 
From October 2012 to September 2020 a $3056 \times 3056$ Kodak KAF-9000 CCD array was in operation for digital image acquisition, with a dynamical range of 16~bit, a pixel size of 12~$\mu$m, and a time resolution of about 150 s. A noise  of 12 e$^{-}$ RMS and a dark current $<1.5$ e$^{-}$ pixel$^{-1}$ s$^{-1}$ characterized that detector.

In order to improve the quality of the images and their time resolution, in September 2020 a new detector was installed at the focal plane of the telescope. A back illuminated sCMOS camera with a sensor of $2048 \times 2048$ pixels, a pixel size of 11~$\mu$m and, a quantum efficiency of $95 \%$ revamped the capabilities of the telescope. In particular, the new fast acquisition capability of this device allows following the dynamic solar processes occurring in the chromosphere in near real time. We changed the default acquisition rate in the center of the H$\alpha$ line from one image per 10 minute to one image per minute. Moreover, with a maximum frame rate of 24 and 48~fps at 12 and 16~bit, respectively, new applications and new restoring techniques can be used to improve the image quality \citep[e.g., using the lucky imaging technique; see][]{Law06, Mackay13}.

We also modified the system by introducing an electronic shutter. The use of an on-sensor rolling shutter overcomes the need for mechanical shutters. This avoids the exposure gradient effects associated with those induced by, e.g, an iris shutter, thus providing much better accuracy for photometry. Moreover, the rolling shutter allows using shorter time exposures and adapting them on the basis of the sky conditions, as well as on the level of solar activity. 

The above mentioned upgrades allowed us to reduce the exposure time from 0.2~s, used with the old setup, down to 0.004~s currently operated.  

An example of an image taken during good seeing conditions can be seen in Figure~\ref{fig:example}.

\subsection{Software Upgrade}
\label{SWup} 

\begin{figure}   
	\centerline{\includegraphics[width=\textwidth,trim= 0 0 0 0,clip]{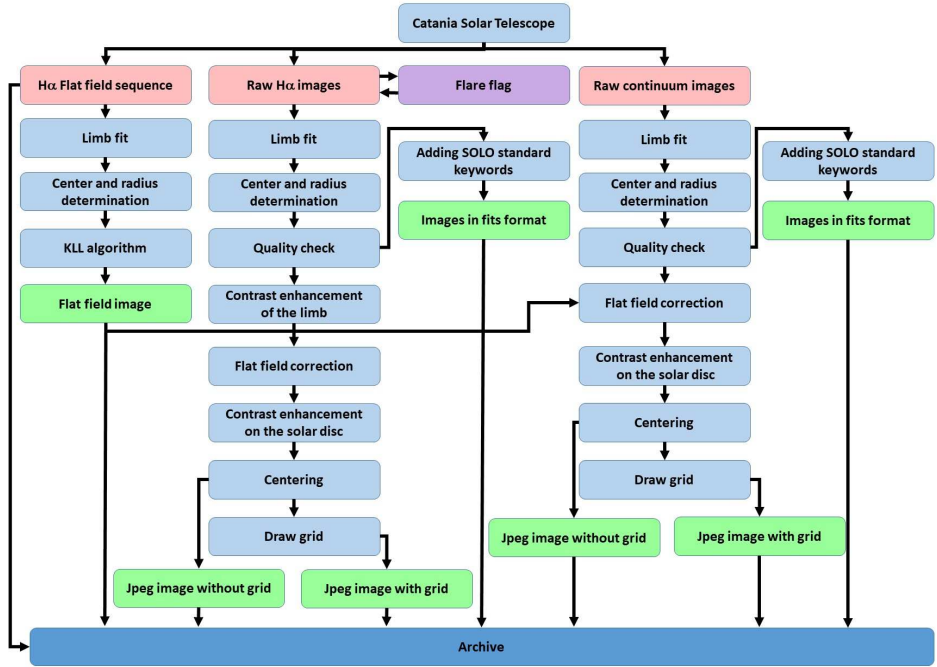}
	}
	\caption{Flow chart describing the new acquisition software operating at INAF-Catania Solar Telescope.}
	\label{fig:flow}
\end{figure}

In addition, an upgrade of the pipeline for the acquisition and handling of the data has also been carried out. This aspect allowed us to improve the quality of the raw images and corrected images. By the new acquisition system, raw images are first transferred to the machine dedicated to the preliminary analysis and data reduction, afterwards they are moved to the storage server (Figure \ref{fig:flow}). 

The raw images (red boxes in Figure \ref{fig:flow}) acquired by the telescope are of three types: H$\alpha$ images taken in order to obtain a flat field image by the KLL method \citep{Kuhn91}, H$\alpha$, and continuum scientific images for observations of the chromosphere and photosphere, respectively.

An H$\alpha$ sequence for flat field purpose is taken every day in about one minute at the beginning of the observing run. The sequence is formed by nine images: one image taken with the center of the solar disk close to the center of the detector, four images with the solar disk almost tangent to each one of the four sides of the detector, and four images with the solar disk near each one of the four corners of the detector. These images are obtained by changing the pointing of the telescope. Assuming that the Sun is a constant light source in one minute, the KLL algorithm uses each pair of pixels illuminated by the same part of the Sun in successive displaced pictures to measure the ratio of the gain of those pixels. By an iteration process it is possible to reconstruct the pixel-to-pixel non-uniformity in the gain of the whole detector \citep[e.g.][]{Li21}.
Therefore, to determine the center and radius of the Sun in the recorded images and to derive the relative displacements among the Sun images a fitting of the solar limb is performed by an IDL routine. After the application of the KLL algorithm the output flat field image, together with the input sequence, are stored in our archive.

By default, every day when weather conditions permit, from 8:00 CET to 13:30 CET, images in the line center of H$\alpha$ are acquired at a cadence of one image per minute. However, a higher cadence (up to one image per second) is taken when a flare is observed in the chromosphere. At the moment the switch to the faster acquisition procedure is performed manually by the observer, however an automatic detection algorithm for flare occurrence based on the near real-time analysis of the H$\alpha$ signal above the solar disk is under development and will be released in the near future (see the violet box in Figure \ref{fig:flow}). 

After the limb fit and the determination of the center and radius of the solar disk, a quality check procedure is applied to the image. A quality index from 1 to 3 (1=good, 3 = bad) is assigned to each image based on the reliability of the solar radius estimation and on the presence of some anomalies in the measured signal. For instance, when a cloud passes in front of the solar disk a quality index equal to 2 or 3 is assigned to the acquired image depending on the significance of the anomalous decrease of the signal. Regardless of the assigned quality index, further keywords are added to the preliminary header, which is initially generated by the acquisition program. These keywords are in agreement with the metadata definition for Solar Orbiter science data \citep{DeGroof19}. The output FITS format data are stored in the archive while the database is updated, accordingly. Indeed, a Ruby script is designed to ease the process of populating the database with the data records. The database containing the records of the images and metadata has been created in MySQL. The same steps are applied also to the continuum images.

Only images of quality class 1 (good) are processed further and additional jpeg images are produced. For the data taken in the center of the H$\alpha$ line the contrast of the limb structures (e.g. solar prominences) is enhanced in order to increase the visibility of their morphology. Then, both chromospheric and photospheric images are corrected by the flat field over the solar disk and an additonal contrast enhancement is performed. In the JPEG2000 images (2048 $\times$ 2048 pixels) the solar disk is centered and institutional logos and a time stamp is added on top. The pipeline produces two versions of JPEG2000 files: with and without the heliographic grid plotted over the solar disk (the former in red scale for the H$\alpha$ images).  

Only images of quality class 1 are available in the ESA Space Weather Service Network, while all images are available in our database\\ (http://ssa.oact.inaf.it/oact/index.html).

\section{Observation of a B5-Class Solar Flare}
	\label{observation} 

Solar flares involve different layers of the solar atmosphere and cover a wide range of energy levels and sizes. This allows us to investigate the physical processes related to the release of energy at different scales. However, the study of small flares has some advantages compared to major events, as flares characterized by a weaker emission do not saturate the digital images and allow analyzing the topology of the involved magnetic systems, like the flare loop configuration or the shape of ribbons, in more detail. Moreover, small flares can be observed entirely within the field of view of high-resolution instruments. However, their unpredictable character makes it difficult to acquire a good quality dataset.

As a showcase of the observational capabilities of the revamped Catania Solar Telescope, we studied a small flare of B5.4 GOES class, that occurred in active region (AR) NOAA~12790 (hereafter, AR~12790) on 7 December 2020 with a peak at 09:55~UT. We used data acquired by the Catania Solar Telescope and by satellite instruments, such as the Interface Region Imaging Spectrograph \citep[\textit{IRIS}:][]{DePontieu14}, and the Helioseismic and Magnetic Imager \citep[HMI:][]{Scherrer12} and Atmospheric Imaging Assembly \citep[AIA:][]{Lemen12} on board the Solar Dynamics Observatory \citep[SDO:][]{Pesnell12}.

\subsection{Dataset Description}
\label{dataset}

We used a sequence of  images taken in the center of the H$\alpha$ line at 656.28 nm by the Catania Solar Telescope on 7 December from 9:30~UT to 10:30~UT. These images with a time cadence of 1~min were characterized by a spatial resolution of about 2\arcsec{} due to the average seeing conditions during the acquisition of that image sequence. Only four images have been neglected in our analysis for their low quality due to clouds in front of the solar disk between 10:01~UT and 10:05~UT.

We also analyzed a simultanoeus observing sequence acquired by IRIS satellite between 09:25:47~UT and 10:14:52~UT on 7 December. The sequence consists of a single large dense 320-step raster scan (OBS3610108077). The sequence had a 0.33\arcsec{} step size and a 9.2~s step cadence, with a pixel size of 0.35\arcsec{} along the \textit{y} direction (spatial binned data), covering a field of view (FOV) of $112\arcsec \times 175\arcsec$. Simultaneously, slit-jaw images (SJIs) were acquired in the 1400, 1330, and 2796~\AA{} passbands, corresponding to the \ion{Si}{IV} 1402~\AA{}, \ion{C}{II} 1335~\AA{} and \ion{Mg}{II}~k lines, respectively. These SJIs have a cadence of 37~s for consecutive frames in each passband and cover a FOV of $167\arcsec \times 175\arcsec$.

To determine the magnetic context of the flaring active region, we used photospheric observations from the SDO satellite consisting of full-disk continuum filtergrams and line-of-sight (LOS) magnetograms taken by HMI along the \ion{Fe}{I}~6173~\AA{} line, with a spatial resolution of 1\arcsec{}. Furthermore, we took advantage of the coronal images acquired by AIA through the EUV filters centered at 131~\AA{}, 193~\AA{}, 171~\AA{}, and 304~\AA{} and UV filters at 1600~\AA{} and 1700~\AA{}. These EUV/UV data have an image spatial scale of about 0.6\arcsec{} per pixel and a cadence of 12~s. 

\begin{figure}   
	\centerline{\includegraphics[width=\textwidth,trim= 0 100 0 20,clip]{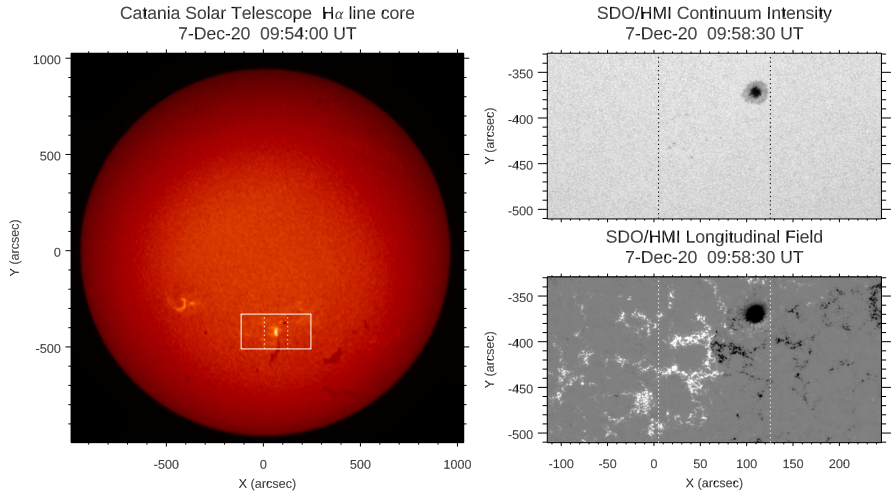}
	}
	\caption{Context images for the Catania Solar Telescope observations relevant to AR~12790. Left panel: Full-disk H$\alpha$ image acquired with the Catania Solar Telescope during the B5.4 flare. The solid box frames the FOV analyzed in the right panels. Right panels: Maps of SDO/HMI continuum intensity (top) and magnetogram (bottom) for the FOV indicated with a solid box in the full-disk H$\alpha$ image. The dashed lines enclose the subFOV further investigated in the article and shown in the following figures.
	}
	\label{fig:context}
\end{figure}

The alignment between the H$\alpha$ images acquired with the Catania Solar Telescope and those acquired by the IRIS and SDO satellites was obtained by applying the IDL SolarSoft mapping routines and cross-correlation techniques with respect to the cospatial subFOV between the different instruments (see Figure~\ref{fig:context}). We used the leading spot of AR~12790 within the subFOV as a fiducial point, taking into account the pixel scale of the different instruments. The accuracy of the alignment is $\pm 1\arcsec$, being comparable to the spatial resolution of SDO/HMI data.

\subsection{Results}
\label{results}	

\begin{table}
	\caption{Flaring events occurring in AR~12790, belonging to the sequence of homologous precursor flares described in the main text. The event observed by the Catania Solar Telescope is indicated using bold letters. 
	}
	\label{tab:flares}
	\begin{tabular}{cccc}     % define the column alignment
		% l: left, c: center, r: right
		\hline                   % horizontal line
		Event & Date & Hour & Flare \\
		&      & peak & class \\
		\hline
		1 & 06--Dec--20 & 23:42~UT  & B8.3 \\
		\textbf{2} & \textbf{07--Dec--20} & \textbf{09:55~UT}  & \textbf{B5.4} \\
		3 & 07--Dec--20 & 14:38~UT  & B6.0 \\
		4 & 07--Dec--20 & 16:32~UT  & C7.4 \\
		\hline
	\end{tabular}
\end{table}

\begin{figure}[t]  
	\centerline{\includegraphics[width=\textwidth,trim= 0 40 0 20,clip]{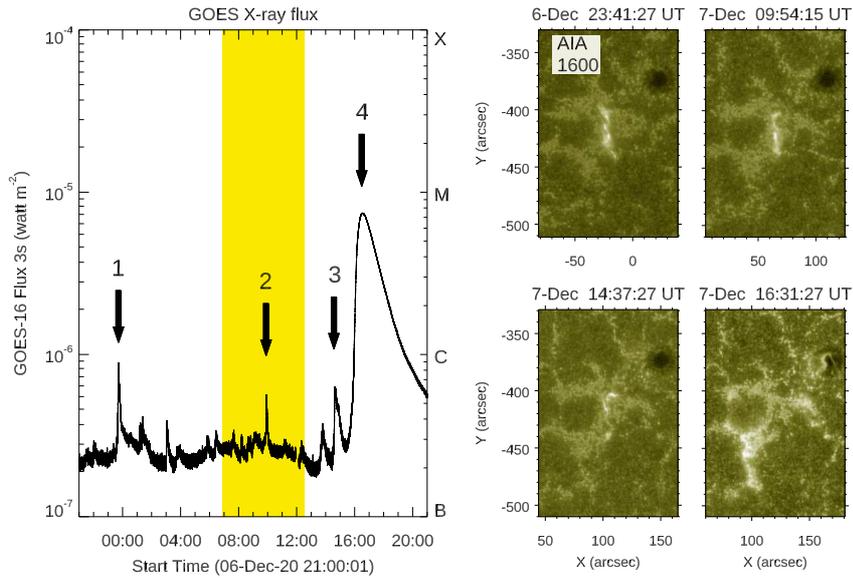}
	}
	\caption{Left panel: Plot of the GOES-16 X-ray flux in the $1-8$~\AA{} passband from 6 December 2020 at 21:00~UT until 7 December 2020 at 21:00~UT. The yellow-shaded area indicates the time interval of the Catania Solar Telescope observations. Events listed in Table~\ref{tab:flares} are indicated with arrows and numbers. Right panels: SDO/AIA 1600~\AA{} filtergrams relative to the peak of the flares belonging to the sequence listed in Table~\ref{tab:flares}.
	}
	\label{fig:sequence_homologous}
\end{figure}

The analyzed flaring event (i.e. SOL2020-12-07T09:55) was part of a sequence of homologous precursor flares \citep[see][]{Romano15} occurring in AR~12790, culminating in a C7.4 flare with a peak at 16:32~UT on the same day. A list of these events is provided in Table~\ref{tab:flares}. The X-ray flux measured by the GOES-16 satellite between 6 December at 21:00~UT and 7 December at 21:00~UT is shown in Figure~\ref{fig:sequence_homologous} (left panel). A sequence of images acquired with the AIA UV filter at 1600~\AA{}, which are relevant to the precursor flares, including the SOL2020-12-07T09:55 event, together with the stronger C7-class flare, is shown in Figure~\ref{fig:sequence_homologous} (right panels). Comparing the location of the flare to the magnetic configuration of AR~12790 reported in Figure~\ref{fig:context}, we can see that the flaring emission during the precursor events takes place along the PIL of AR~12790, with a compact ribbon aligned along the north-south direction. As we can infer from the comparison between the H$\alpha$ image and SDO/HMI magnetogram, a filament characterized by an elongated S-shape is located along the PIL. The northern and southern ends of this filament correspond to the eastern side (towards negative $x$ axis) of the preceding sunspot of the AR and the southern magnetic field concentrations of the AR, respectively (see Figure~\ref{fig:context}). The location of the flare ribbons suggests that only a portion of the filament is involved by the recurrent flares, which are not able to completely destabilize the main magnetic field configuration supporting the filament.

\begin{figure}   
	\centerline{\includegraphics[width=\textwidth,trim= 0 40 0 0,clip]{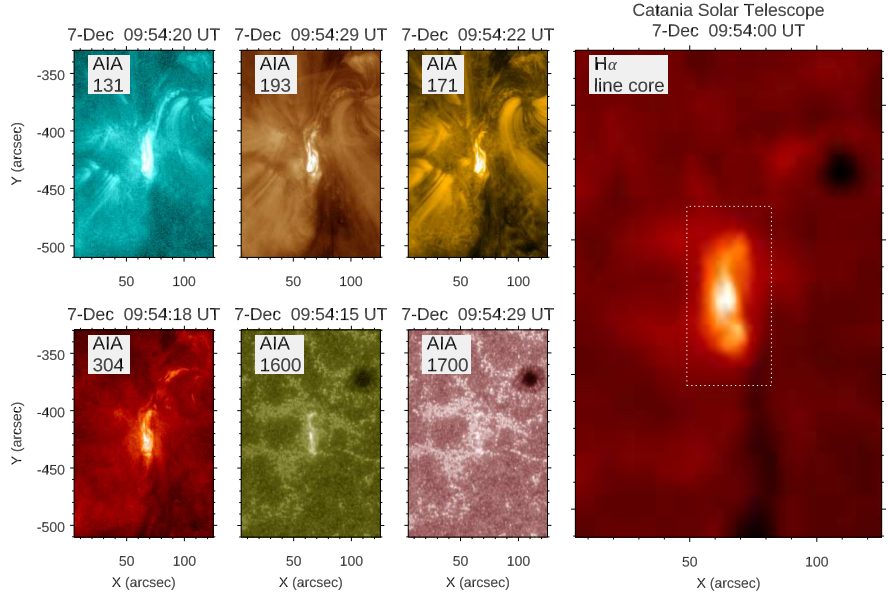}
	}
	\caption{Simultaneous multi-wavelength observations during the peak of the SOL2020-12-07T09:55 flare. Left panels: Maps derived from SDO/AIA observations corresponding to filters with decreasing temperature formation, from EUV 131~\AA{} down to UV 1700~\AA{}. Right panel: H$\alpha$ observation acquired with the Catania Solar Telescope. The dashed box in the H$\alpha$ map indicates the subFOV used for computing the light curve shown in Figure~\ref{fig:light_curve} and for studying the sequence of IRIS observations shown in Figure~\ref{fig:sequence_iris}.
	}
	\label{fig:multi_flare}
\end{figure}

Figure~\ref{fig:multi_flare} displays simultaneous multi-wavelength observations acquired at the peak of the SOL2020-12-07T09:55 event, during the time interval covered by the Catania Solar Telescope observations. A compact ribbon is observed at decreasing temperature formation heights, according to the response of AIA EUV and UV filters, taking into account emission contributions due to flares \citep{ODwyer10}. At 304~\AA{} a bright patch protruding from the ribbon is observed near the preceding sunspot of AR~12790. This emitting structure is also faintly visible in the H$\alpha$ map (see the right panel of Figure~\ref{fig:multi_flare}), and it is clearly detected in the simultaneous \textit{IRIS} SJIs at 1400~\AA{} (not shown here). Probably this structure which is visible mainly at a chromospheric level can be interpreted as the emission produced in the northern footpoint of the filament by the electron beams, characterized by particular high energy and reaching the lower layers of the solar atmosphere. For this reason, probably, this bright region is not visible in the corona by AIA EUV filters.

\begin{figure}   
	\centerline{\includegraphics[width=\textwidth,trim= 0 30 0 100,clip]{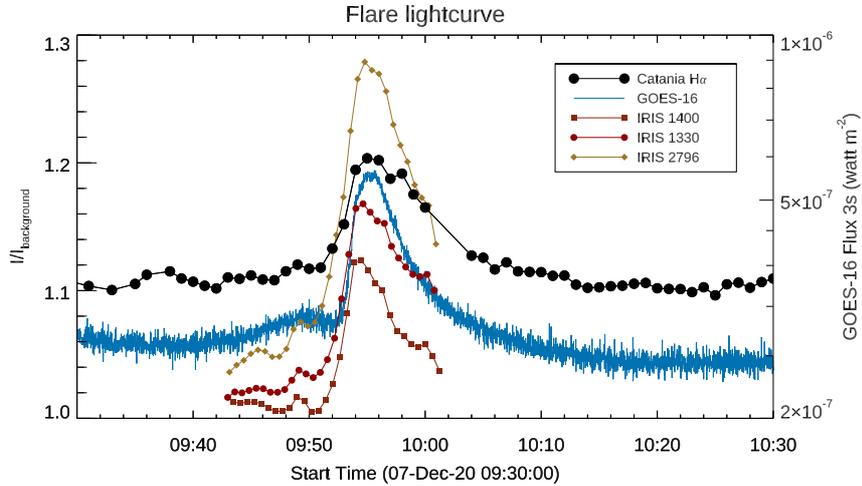}
	}
	\caption{Lightcurves computed from simultaneous multi-wavelength observations during the SOL2020-12-07T09:55 flare: H$\alpha$ data acquired with the Catania Solar Telescope (black symbols) and IRIS SJIs at 1400, 1330, and 2796~\AA{} (see the legend), together with the GOES-16 X-ray flux in the $1-8$~\AA{}. All the curves are referring to the subFOV indicated with a dashed box in the H$\alpha$ map in Figure~\ref{fig:multi_flare}.
	}
	\label{fig:light_curve}
\end{figure}

The subFOV indicated with a dashed box in the H$\alpha$ map of Figure~\ref{fig:multi_flare} has been used to compute the light curves of the SOL2020-12-07T09:55 flare in different wavelengths. Figure~\ref{fig:light_curve} shows the light curve deduced from the H$\alpha$ sequence (black symbols), as well as the light curve relevant to the IRIS SJIs used in this work (1400, 1330, and 2976~\AA{}, colored symbols). The H$\alpha$ light curve has been computed by dividing the average value of the brightness in the subFOV by the brightness value at the disk center, in order to get rid of global observational effects, such as the presence of clouds. Therefore, a value larger than 1 at the beginning of the light curve indicates that we were observing a facular region. The background value for IRIS has been set by considering the average value in the subFOV at the beginning of SJI observations. For ease of comparison, we also plot in the same graph the GOES-16 X-ray flux in the $1-8$~\AA{} (blue line). Despite the small intensity of the flare (B5.4 GOES class), we remark that the sensitivity of the Catania Solar Telescope is able to detect the increase of the brightness in the center of the H$\alpha$ line and to highlight the almost contemporary occurrence of the light curve peak in comparison to the IRIS wavelengths.

\begin{figure}   
	\centerline{\includegraphics[width=\textwidth,trim= 0 30 0 100,clip]{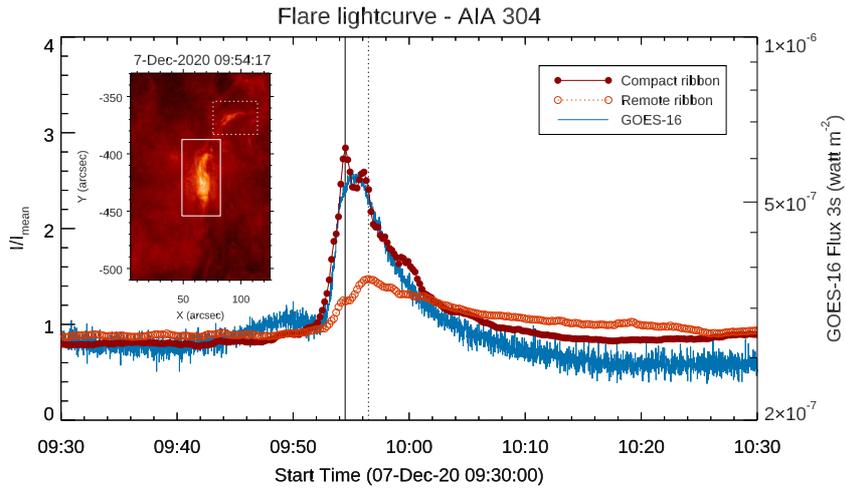}
	}
	\caption{Lightcurves computed from AIA 304~\AA{} filtergrams, relevant to the boxes indicated in the inset. Note that the solid box, framing the compact ribbon, encloses the same subFOV indicated with a dashed box in the H$\alpha$ map as in Figure~\ref{fig:multi_flare}. The dashed box encloses the remote ribbon. For comparison, we include the GOES-16 X-ray flux in the $1-8$~\AA{}. The vertical continuum and dashed lines indicate the times corresponding to the emission peak of the main and northern flare ribbons, respectively.
	}
	\label{fig:light_curve_304}
\end{figure}

In Figure~\ref{fig:light_curve_304} we illustrate the time delay between the intensity peak occurring in the lightcurves relative to the SDO/AIA 304~\AA{} channel for the compact ribbon and the remote ribbon of the flare, respectively.
This time delay is 120~s, indicating a significant travel time for the propagation of the disturbance in comparison to a classical spine configuration \citep{Masson12}.

\begin{figure}  
	\centerline{\includegraphics[width=0.85\textheight,trim= -10 40 0 -10,clip,angle=90]{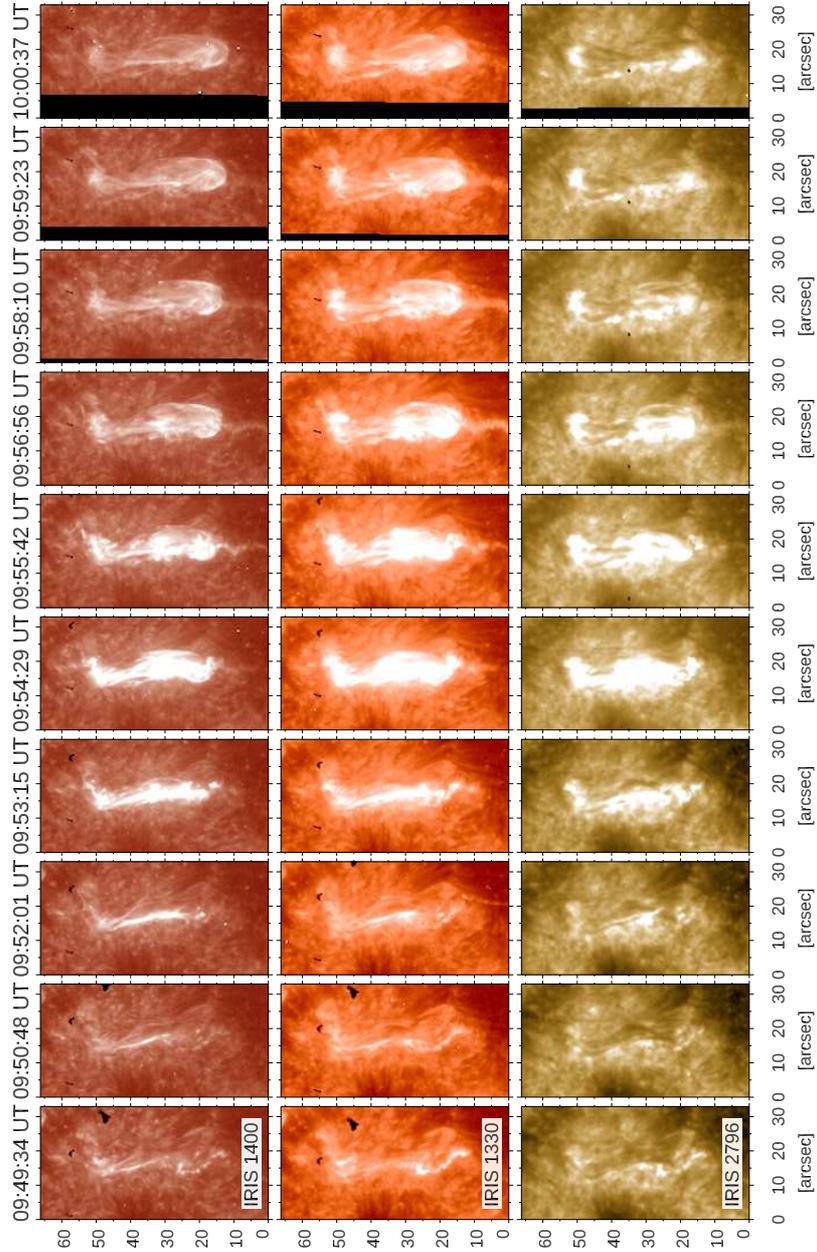} 
	}
	\caption{Sequence of IRIS SJIs at 1400, 1330, and 2796~\AA{} acquired during the development of the SOL2020-12-07T09:55 flare, relative to the subFOV indicated with a dashed box in the H$\alpha$ map in Figure~\ref{fig:multi_flare}. All the images for the same passband are rescaled to the same minimum and maximum values. 
	}
	\label{fig:sequence_iris}
\end{figure}

Finally, in Figure~\ref{fig:sequence_iris} we report the observations relative to the IRIS SJIs for the restricted subFOV indicated in Figure~\ref{fig:multi_flare}. These high resolution images allow us to obtain a better insight about the fine structure of the compact ribbon of the SOL2020-12-07T09:55 flare. 

Indeed, the compact ribbon appears to consist of threads, especially in its northern part. At the beginning of the sequence (09:49~UT) in the \ion{Mg}{II}~k (2796 \AA) images, the dark filament shows its twisted fine structure, while its counterpart is very thin and bright in the 1400 and 1330~\AA{} passbands. After the peak, when all the IRIS SJIs are saturated and the compact ribbon is observed, it is possible to detect a bifurcation of the filament, especially in the \ion{C}{II} (1330 \AA) and \ion{Mg}{II} k images. This behaviour is typical for the homologous flares characterized by the reformation of the involved filament after each event by the splitting of a single flux rope during the eruption \citep[e.g.][]{Gibson06} or by the eruption of the upper part of a double-decker system \citep[e.g.][]{Kliem14}.

\section{Discussion and Conclusions}
	\label{Discussion} 	

%One of the principal manifestations of the flare energy release is the presence of the ribbon brightenings in chromosphere, that are after observed in the ultraviolet wavelenghts. They are the effects of the accelerated particles from the reconnection site to the lower layers of the atmosphere.
%The location, the shape, and the motion of the ribbons provide useful information to understand the connectivity of the magnetic systems involved by the flare. Sometimes the ribbons show a peculiar shape which suggest some idea of the overlying topology of the magnetic field and an indication of the main reconnection site \citep[see, e.g.,][]{Sav15}. For example, circular ribbons are often interpreted as the evidence of the presence in corona of a three-dimensional null point \citep{Mas09, Rei12, Wan12, Sun13, Jia14, Man14, Zha15, Liu15}. Such a topological field model is associated to many flares, although it is not easy to observe this kind of ribbons, probably due to the asymmetry of the real configuration and to the privileged direction of particle acceleration. The presence of a 3D null point determines the formation of a surface, named fan, and two singular field lines, named inner and outer spine, belonging to two different connectivity domains \citep{Mas12}. Both the fan and the spines are sites where current sheets form.
%However, the conditions which lead to the occurrence of solar flares in such configuration have been rarely studied so far due to the difficulties to observe these topologies with enough spatial resolution.
In this article we have analyzed observations of a GOES B5-class flare occurring in a moderately complex active region and producing a compact ribbon at the chromospheric level and a farther remote ribbon. This event provides a contribution to the few observations of such a kind of multiple-ribbon events reported in the literature, which can be interpreted in the light of models that invoke full 3D slipping-reconnection and elongated ribbons \citep{Pontin16}. 

The good observing performance of the revamped Catania Solar Telescope has allowed us to investigate the evolution of the H$\alpha$ ribbon of the SOL2020-12-07T09:55 flaring event. Despite the low intensity of the flare which was of B5.4 GOES level, the H$\alpha$ images showed the sensitivity of Catania Solar Telescope to detect the variation of emission in time along the chromospheric ribbons, with a good reliability.

Indeed, the presence of the ribbon brightening in the chromosphere, which is also observed in the ultraviolet wavelengths, is one of the principal manifestations of the flare energy release in the solar atmosphere. Using the Catania Solar Telescope and space-based instrument data we were able to determine that also this faint flare was characterized by the splitting of a single flux rope during the eruption as usually observed in stronger events. Moreover, the temporal delay of the brightening observed in the chromosphere between the region along the PIL and the region corresponding to the northern footpoint of the filament allows us to infer the presence of a mechanism able to transfer the instability from the flux rope forming the main body of the filament to the northern footpoint of the filament. Although the emission produced in the northern footpoint of the filament is only a few arcseconds away from the main ribbon, the reported delay of about 120~s between the peaks could be attributed to the twisted configuration of the magnetic field of a flux rope. In fact, the helical pattern of the flux rope, detected in the EUV images, could justify a longer time for the propagation of the signal in comparison to the typical time employed for the acceleration of the particles along a spine in a 3D null point configuration \citep{Romano17}.

This showcase clearly demonstrates the added value of the observations of the Catania Solar Telescope, which are useful both for space weather forecasting purpose and for scientific exploitation.
In the near future, we plan to further upgrade the quality of the service provided by this telescope to the space weather community by means of the implementation of techniques which improve the data quality, e.g. the lucky imaging technique, and an automatic detection algorithm for flare occurrences.

\begin{acks}
This research received funding from the European Union’s Horizon 2020 Research and
Innovation 531 program under grant agreements No 824135 (SOLARNET) and support by the Italian Space
Agency (ASI) under contract 2021-12-HH.0 to the co-financing INAF for the Italian contribution to the
Solar-C EUVST preparatory science programme. This work was supported by the Italian MIUR-PRIN 2017
on “Space Weather: impact on circumterrestrial environment of solar activity” and by the Universitá degli
Studi di Catania (Piano per la Ricerca Universitá di Catania - Linea di intervento 2 “PIACERI”).
\end{acks}

\section*{Declarations}
{\bf Disclosure of Potential Conflicts of Interest} The authors declare that there are no conflicts of interest.

\end{article} 

\end{document}